\documentstyle[12pt,aps,floats,prc,epsfig,tighten]{revtex}
\def\Journal#1#2#3#4{{\em #1} {\bf #2}, #3 (#4) }
\def\NPA{{ Nucl. Phys.} A}

\def\PREP{ Phys. Rep.}
\def\PRC{{Phys. Rev.} C}
\def\PL {Phys. Lett.}
\newcommand{\matrixel}[3]{\mbox{$\left<#1|#2|#3\right>$}}

\newcommand{\symprime}[1]{{#1}^{\prime}}

\begin{document} 
\title{ $\Delta(1232)$ Isobar Excitations in Nuclear
Many-Body Systems derived from various NN Interactions}
\author{T. Frick, S. Kaiser,  H. M\"uther}
\address{Institut f\"ur
Theoretische Physik, \\ Universit\"at T\"ubingen, D-72076 T\"ubingen, Germany}
\author{ A. Polls} 
\address{Departament d'Estructura i Constituents de la Mat\`eria,\\ 
Universitat de Barcelona, E-08028 Barcelona, Spain}
\maketitle

\begin{abstract}
$\Delta$ isobar components in the nuclear many-body wave function are
investigated for the deuteron, light nuclei (${}^{16}\mbox{O}$) and infinite 
nuclear matter within the framework of the coupled-cluster theory. The
predictions derived for various realistic models of the baryon-baryon
interaction are compared to each other. These include local ($V28$) and 
non-local meson exchange potentials (Bonn$_{2000}$) but also a model recently
derived by the Salamanca group accounting for quark degrees of freedom. The
characteristic differences which are obtained for the $N\Delta$ and
$\Delta\Delta$ correlation functions are related to the approximation made in
deriving the matrix elements for the baryon-baryon interaction.
\end{abstract}
\section{Introduction\label{Inroduction}}

It has always been one of the basic challenges of theoretical nuclear physics to
derive the basic properties of nuclei from realistic models of the
nucleon-nucleon (NN) interactions. Very sophisticated approximation schemes 
have been developed to solve the many-body problem of strongly interacting
nucleons. As examples we mention Brueckner hole-line expansion\cite{jeuk}, the
coupled-cluster or ``exponential S'' approach\cite{kuem}, the self-consistent
evaluation of Greens functions\cite{wim}, variational approaches using
correlated basis functions\cite{fhnc} and recent developments employing quantum
Monte-Carlo techniques\cite{monc}. (see e.g.~the recent review \cite{mupo} on
these many-body approaches).

A major ingredient for such investigations is the definition of a realistic NN
interaction. Here we define a realistic NN interaction to be a model for the
nucleon nucleon interaction in which the parameters have been adjusted to
obtain a very good fit of the experimental NN scattering data at energies up to
the threshold for $\pi$ production. As examples for modern realistic
interactions we refer to refs.\cite{cdb,argv18,nijm1}. 

The central assumption of such investigations is that the nucleons can be
considered as inert particles interacting by two-body forces, which are
identical in the vacuum and in the nuclear medium. It is of course well known
that nucleons are not elementary particles. 
At short distances, nucleons will polarize each other, which will lead 
to virtual excitations. These sub-nucleonic degrees of freedom are
effectively taken into account in the realistic NN interaction as the parameters
are adjusted to the data.

As an example we mention the processes of the type that two nucleons interact
with each other exchanging a $\pi$ or $\rho$ meson leading to a $N\Delta$ or
$\Delta\Delta$ state. After the exchange of a second meson the two baryons may
be returned into a $NN$ state such that this process contributes to the
amplitude for elastic $NN$ scattering. A theoretical approach which accounts for
interacting nucleons only parameterizes this process with intermediate isobar
excitations by adjusting e.g.~the coupling constant of the $\sigma$ meson in a
One-Boson-Exchange model for the $NN$ interaction\cite{elster}.    
If, however, the isobar configurations are included explicitely one finds that
a large part of the medium range attraction in the $NN$ interaction originates 
from the coupling to the $N\Delta$ and $\Delta\Delta$ configurations.

Both kinds of models, with and without the explicit consideration of isobar
configurations, yield essentially the interaction for two nucleons in the
vacuum, since they are adjusted to reproduce the $NN$ phase shifts. Remarkable
differences show up, however, if the different kinds of interactions are
considered for the interaction of two nucleons inside a nuclear medium. While
the exchange of the $\sigma$ meson is essentially the same in the vacuum and in
the nuclear medium, the formation of virtual $N\Delta$ and $\Delta\Delta$
configurations is suppressed in nuclear matter (due to Pauli effects and binding
of the nucleons) as compared to the vacuum. This leads to a loss of attraction
for the effective interaction of two nucleons in the medium, which results in a
smaller calculated binding energy. This effect has been investigated by various
groups using the Brueckner-Hartree-Fock approximation, which means that
correlations between nucleons are taken into account by means of the effective
interaction, the $G$-matrix\cite{green,anast,manz}.

The isobar degrees of freedom may of course also be considered explicitely in
the solution of the nuclear many-body problem, i.e. one allows for many-body wave
functions which contain isobar configurations. Such investigations have been
made e.g.~for the three-nucleon problem\cite{hajduk} and for nuclear
matter\cite{manz}. Presently, there is a renewed interest in the study of these
isobar configurations in the nuclear wave function. One reason is the attempt
to explore the isobar components in the nuclear wave function by means of
different experiments. This includes the photoproduction of pions\cite{huber}
and the isobar current contribution to exclusive ($e,e'NN$) reactions. Another
reason is the development of new models for the baryon-baryon interaction
including isobar configurations\cite{V28,entem,mac00}.

In this study we would like to investigate the predictions for the isobar
components in the nuclear wave functions derived from various models
for the baryon-baryon interaction. We are going to compare the isobar amplitudes
calculated for the Argonne $V28$ potential\cite{V28} with those derived   
from a recent update\cite{mac00} of Model I in table B.1 of \cite{rupr}, which
we will denote as Bonn$_{2000}$. Both of these models are based on the meson
exchange picture of the baryon-baryon interaction. This means that e.g.\ the
transition potential $NN \to N\Delta$ is dominated by the contribution from 
$\pi$ exchange. While the Bonn$_{2000}$ approach considers the complete
relativistic structure of this term, the $V28$ model reduces this $\pi$ exchange
contribution to a local potential. This $\pi$ exchange contribution is
supplemented by the $\rho$ exchange in the Bonn$_{2000}$ model. The $V28$
approach on the other hand adds a phenomenological contribution of short range.
In \cite{local} it has been observed that the reduction of meson exchange
interactions to the local approximation yields characteristic differences. Can
such differences also be observed for the isobar contributions?

A quite different approach has recently been developed by the Salamanca
group\cite{entem}. They derive a NN interaction in the framework of the Chiral
Quark Cluster (CQC) model. The problem of two interacting clusters (baryons) of
quarks is solved by means of the resonating group method. The Pauli principle
between the interacting quarks is an important source for the short-range
repulsion of the NN interaction\cite{lueb}. At large distances the $\pi$ 
exchange between the quarks in the two clusters evolves to the $\pi$ exchange
between two baryons. At shorter distances, however, this non-local model 
for the baryon-baryon interaction might yield results that are quite different from 
a meson-exchange picture. This Salamanca potential does not give such a perfect fit to the
NN scattering phase shifts as the Bonn$_{2000}$ or the $V28$. For the $^1S_0$
and $^3S_1 - ^3D_1$ partial waves of the NN system, however, the agreement with the
empirical data is rather good.

We are going to compare correlated two-baryon wave functions in the deuteron,
${}^{16}\mbox{O}$ and infinite nuclear matter derived from these three models of the
baryon-baryon interaction. While the deuteron wave functions are obtained from
an exact diagonalization in momentum space, the wave functions for  ${}^{16}\mbox{O}$ 
and nuclear matter are calculated in the framework of the ``exponential S'' 
approach\cite{kuem}, generalized to include isobar degrees of freedom and
evaluated in momentum representation.

After this introduction we discuss some differences between the transition
potentials for $V28$ and  Bonn$_{2000}$ in the next section. The wave functions
of the deuteron derived from these three models are compared in section 3, while
the correlation functions for  ${}^{16}\mbox{O}$ and infinite nuclear matter are
presented in section 4. A summary and conclusions are added in the final
section. 

\section{Transition potentials in different models}
\label{Transition potentials in different models}
The key for an understanding of isobar components in the many-body
wave function of nuclear systems is the transition potential, i.e.~the matrix
elements of the baryon-baryon interaction connecting two-nucleon states to
$N\Delta$ and $\Delta\Delta$ configurations.
A transition $\mbox{N}\leftrightarrow\Delta$ implies the change of 
both spin and isospin from $\frac{1}{2}$ to $\frac{3}{2}$. As isospin 
conservation has to be fulfilled at each vertex, such a transition requires
the emission or absorption of an isovector meson, if one is considering a 
meson-exchange model for the baryon-baryon interaction. The most important 
isovector meson contributing to such transitions is of course the pion.
The $\rho(770)$ has a considerable effect, too, but mainly
for large momentum transfers between the interacting baryons.

In this section we would like to compare the one-pion-exchange (OPE)
contribution as it is included in the Bonn$_{2000}$ interaction with the
corresponding term in the Argonne $V28$ interaction. As an example we consider
the transition potential $NN \leftrightarrow\Delta\Delta$, in particular we will
focus the attention to the central component as it shows up in $^3S_1$ partial
waves. This OPE  transition amplitude has been 
formulated in a helicity representation in which the quantisation axis of the 
particle spins are the directions of the relative momenta 
$\mathbf{q}$ and $\mathbf{q^{\prime}}$ for the initial and final state,
respectively \cite{hol77}:  
\begin{eqnarray}
\matrixel{{\mathbf{q^{\prime}}}\,\Lambda_{N_1}\Lambda_{N_2}}
{V^{\pi}}
{{\mathbf{q}}\,\Lambda_{\Delta_1}\Lambda_{\Delta_2}} 
&=&
\frac{4\pi}{(2\pi)^3}\,
\frac{f_{N\Delta\pi}^{2}}{m_{\pi}^2}\,
c(T)\,
(q^{\prime}_{\mu}-q_{\mu})(q^{\prime}_{\nu}-q_{\nu})\,
F_{\pi}^2({\mathbf{q'}}-{\mathbf{q}})\nonumber \nonumber
\\ 
&&\quad \times\frac
{
\bar{u}(-{\mathbf{q^{\prime}}},{\Lambda_{N_2}})\,
u^{\mu}(-{\mathbf{q}},{\Lambda_{\Delta_2}})\,
\bar{u}({\mathbf{q^{\prime}}},{\Lambda_{N_1}})\,
u^{\nu}({\mathbf{q}},{\Lambda_{\Delta_1}})}
{\sqrt{({\mathbf{q'}}-{\mathbf{q}})^2+m_{\pi}^2}
\left[\sqrt{({\mathbf{q'}}-{\mathbf{q}})^2+m_{\pi}^2}+m_{\Delta}-m_N\right]} 
\label{feyn} 
\mbox{.} 
\end{eqnarray}
$c(T)$ is the isospin factor that remains from isospin matrix elements and
corresponds to $c(T)= -\sqrt{2}$ in this channel. The energy denominator in
this expression contains the masses of the pion $m_\pi$, the nucleon $m_N$ and
the $m_\Delta$. It deviates from a simple pion-exchange propagator, as it 
has been derived from time-dependent perturbation theory accounting for the mass
difference between $m_N$ and $m_\Delta$\cite{dur77}. The function $F_\pi$
represents the form factor for each $\pi N\Delta$ vertex. 
The initial relative momentum $q$ and the final relative 
momentum $q^{\prime}$ are chosen along the $z$ axis, $q=(0,0,0,q)$,
and in the $x$-$z$-plane, $q^{\prime}=
(0,q^{\prime}\sin{\theta},0,q^{\prime}\cos{\theta})$, respectively,
where $\theta$ is the angle between $\mathbf{q}$ and $\mathbf{q}^{\prime}$. 
$\Lambda_{B_i}$ labels the helicity quantum numbers of the spinors.
For the $\Delta$ isobars these spinors fulfil the Rarita-Schwinger equations 
 and can be constructed explicitely by coupling
a vector field (spin 1) and a Dirac field (spin $\frac{1}{2}$)
\begin{equation}  
u_{\mu}({\mathbf{q}},\Lambda) = \sum_{\lambda,s}\,
C_{1{\lambda}{\frac{1}{2}}s}^{{\frac{3}{2}}\Lambda}\,
e_{\mu}({\mathbf{q}},\lambda)\,
u({\mathbf{q}},s)
\mbox{,}
\end{equation}
with 
\begin{equation}
\label{rss}
e_{\mu}({\mathbf{q}},\lambda)=
\left(\frac{\hat{{\mathbf{e}}}_{\lambda}\!\cdot\!{\mathbf{q}}}
{m_{\Delta}},
-\hat{{\mathbf{e}}}_{\lambda}-\frac{{\mathbf{q}}(\hat{{\mathbf{e}}}_
{\lambda}\!\cdot\!{\mathbf{q}})}
{m_{\Delta}(\sqrt{{\mathbf{q}}^2+m_{\Delta}^2}+m_{\Delta})}\right)\mbox{,}
\end{equation}
where the $C_{1{\lambda}{\frac{1}{2}}s}^{{\frac{3}{2}}\Lambda}$ are 
Clebsch-Gordan coefficients in the notation of~\cite{var88} 
and $u(\mathbf{p},s)$ is a Dirac spinor in spin state $s$. 
$\hat{\mathbf{e}}_{+}$, $\hat{\mathbf{e}}_{0}$ and $\hat{\mathbf{e}}_{-}$
are the circular polarisation vectors.

From both the Dirac spinors for the nucleons 
and the Rarita-Schwinger spinors for the Deltas, 
the complex momentum structure is now removed by taking their value at 
$\mathbf{q}=\mathbf{q}^{\prime}=0$.
It is straightforward to obtain
the helicity Feynman graph structure (\ref{feyn}) for all 
$2\times2\times4\times4=64$ combinations of helicity projection numbers
in this `static' approximation. 
Most of them can be derived from symmetry 
relations~\cite{hol77}, and one only needs to evaluate 
10 matrix elements explicitely. 

The helicity amplitudes are then projected onto states with definite
total relative angular momentum $J$:
\begin{eqnarray}
\big<{q^{\prime}\,J^{\symprime{\Lambda}}\Lambda_{N_1}\!\Lambda_{N_2}}
\big|{V^{\pi}}\big|
{q\,J^{\Lambda}\Lambda_{\Delta_1}\!\Lambda_{\Delta_2}}\big>
&=&2\pi\!\int_{-1}^1\!
{\rm{d}}(\cos\theta)
d_{\Lambda,\Lambda^{\prime}}^J(\theta)
\big<{{\mathbf{q^{\prime}}}\Lambda_{N_1}\!\Lambda_{N_2}}
\big|{V^{\pi}}\big|
{{\mathbf{q}}\Lambda_{\Delta_1}\!\Lambda_{\Delta_2}}\big>
\mbox{.}
\nonumber\\
\end{eqnarray}
Here, $d^J_{\Lambda,\Lambda^{\prime}}(\theta)$
are the reduced rotation matrices
with $\Lambda=\Lambda_{\Delta_1}-\Lambda_{\Delta_2}$ and
$\Lambda^{\prime}=\Lambda_{N_1}-\Lambda_{N_2}$.
In the next step one performs the transformation into the basis of the partial
wave states, in which orbital angular momentum $L$ and spin $S$ are coupled to 
$J$:
\\
\begin{eqnarray}
\big<{q^{\prime}\,{}^{2\symprime{S}+1}\!L^{\prime NN}_J}
\big|{V^{\pi}}\big|
{{q}\,{}^{2S+1}L^{\Delta\Delta}_J}\big>  
& = &
\frac{\sqrt{(2\symprime{L}+1)(2L+1)}}{2J+1}
\!\!\!\!\!\!\!\!\!\!
\sum_{\Lambda_{N_1}\Lambda_{N_2}\Lambda_{\Delta_1}\Lambda_{\Delta_2}}
\!\!\!\!\!\!\!\!\!\!
C_{\symprime{L}0\symprime{S}\symprime{\Lambda}}^{J\symprime{\Lambda}}
C_{\frac{1}{2}\Lambda_{N_1}\frac{1}{2}-\Lambda_{N_2}}
^{\symprime{S}\symprime{\Lambda}}
\nonumber
\\
& & \quad 
\big<q^{\prime}\,{J^{\symprime{\Lambda}}\,\Lambda_{N_1}\Lambda_{N_2}}
\big|{V^{\pi}}\big|
{q\,J^{\Lambda}\,\Lambda_{\Delta_1}\Lambda_{\Delta_2}}\big>
C_{\frac{3}{2}\Lambda_{\Delta_1}\frac{3}{2}-\Lambda_{\Delta_2}}
^{S{\Lambda}}
C_{{L}0{S}{\Lambda}}^{J{\Lambda}}
\mbox{.}
\end{eqnarray}
The transformation coefficients have been taken from~\cite{jac59}.

In order to compare these matrix elements for the transition potential in the
Bonn$_{2000}$ model with the corresponding one for the $V28$, we consider two
approximations:
\begin{itemize}
\item[{\bf --}] In evaluating the helicity amplitudes of eq.(\ref{feyn}) the
non-relativistic static limit is considered, i.e.~the Dirac spinors as well as 
the Rarita-Schwinger are considered for $\mathbf{q}=\mathbf{q}^{\prime}=0$.
\item[{\bf --}] The energy denominator in eq.(\ref{feyn}) is replaced by the
usual pion propagator by ignoring the mass difference $m_\Delta - m_N$.
\end{itemize} 
Using these approximations an analytical expression is obtained for
\begin{eqnarray}
\label{resultloc}
\big<{q^{\prime}\,{}^3\!S_1^{NN}}\big|{V^{\pi}}\big|
{{q}\,{}^3\!S_1^{\Delta\Delta}}\big>
&=&-\frac{4\pi^2}{(2\pi)^3}
\frac{\sqrt{40}}{9}
\frac{f_{N\Delta\pi}^2}{m_{\pi}^2}
\!\int_{-1}^{1}\!d(\cos\theta)\,
\left[
1-\frac{m_{\pi}^2}
{({\mathbf{q'}}-{\mathbf{q}})^2+m_{\pi}^2}
\right]
F_{\pi}^2({\mathbf{q'}}-{\mathbf{q}})
\end{eqnarray}
that depends on the momentum transfer $\mathbf{q'}-\mathbf{q}$ only. This means
that the interaction is local. Transforming this expression for the central part
of the transition potential into configuration space, we can identify a Yukawa
term and a contact term, both multiplied by the form factor. 

This expression can be compared with the local transition potential as defined
for the Argonne $V28$ interaction,
\begin{eqnarray}
\label{argonnesigmatau}
\big<{r\,{}^3\!S_1^{NN}}\big|{V^{\pi}}\big|
{{r}\,{}^3\!S_1^{\Delta\Delta}}\big>
& = &\left(\frac{f_{N\Delta{\pi}}^2m_{\pi}}{12{\pi}}\right)
\frac{e^{-{m_{\pi}}r}}{{m_{\pi}}r}(1-e^{-cr^2})
\big<{{}^3\!S_1^{NN}}\big|({\mathbf{S}}_1{\mathbf{S}}_2)^{\dagger}
({\mathbf{T}}_1{\mathbf{T}}_2)^{\dagger}\big|
{{}^3\!S_1^{\Delta\Delta}}\big>\mbox{,}
\end{eqnarray}   
with $\mathbf{S}_i$ and $\mathbf{T}_i$ denoting the transition operator for spin
and isospin. Note that comparing this expression with the local approximation to
the Bonn$_{2000}$ model in (\ref{resultloc}), the contact term is removed and the
form factor is replaced in $V28$ by a Gaussian cutoff to regularize the 
matrix elements for small interparticle distances.

\begin{figure}[t]
\begin{center}
\epsfig{figure=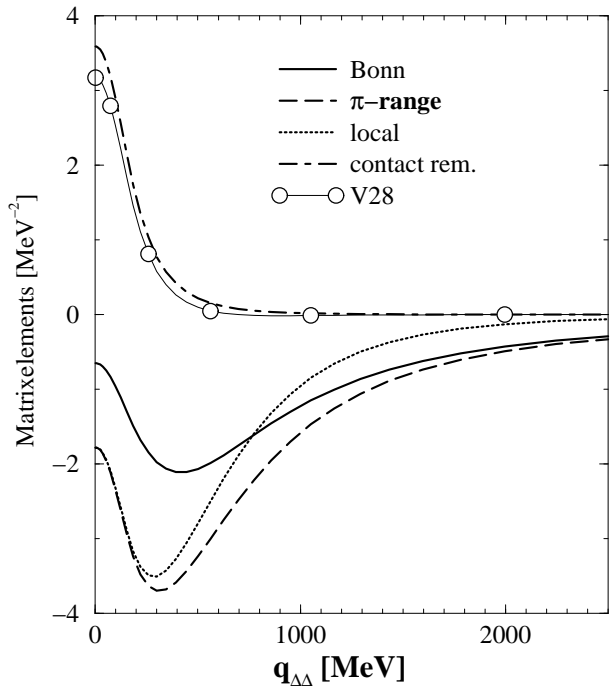,width=8cm}
\end{center}
\caption{Matrix elements for the one-pion-exchange contribution to the
transition potential $\big<{q^{\prime}\,{}^3\!S_1^{NN}}\big|{V^{\pi}}\big|
{{q}\,{}^3\!S_1^{\Delta\Delta}}\big>$ assuming various approximations as
discussed in the text. Results are displayed as a function of the momentum $q$
in the $\Delta\Delta$ state. The momentum of the $NN$ state has been fixed to
$q$ = 100 MeV. All values have been multiplied with a common factor of
$10^6$. \label{fig1}}
\end{figure} 
Matrix elements of the one-pion-exchange (OPE) contribution to the transition
potential in the partial wave
$\big<{q^{\prime}\,{}^3\!S_1^{NN}}\big|{V^{\pi}}\big|
{{q}\,{}^3\!S_1^{\Delta\Delta}}\big>$ for these different approaches are
displayed in Fig.~\ref{fig1}. The relative momentum of the NN pair $q'$ is
fixed at  $100\,\mbox{MeV}$ and the matrix elements are presented as a
function of the relative momentum for the $\Delta\Delta$ state.

The mass difference $m_\Delta - m_N$ in the pion propagator of (\ref{feyn})
has a remarkable effect on the calculated transition potential. This can be
seen in Fig.~\ref{fig1} from the comparison of the solid line (Bonn), which represents the complete
OPE contribution in the Bonn$_{2000}$ potential, with the long dashed line
($\pi$-range), which exhibits the results obtained after replacing the pion
propagator of (\ref{feyn}) by the conventional $\pi$ propagator. The inclusion
of the mass difference $m_\Delta - m_N$ leads to a quenching  of the transition
potential by about a factor two for small momenta, while the two curves
approach each other at large values of $q_{\Delta\Delta}$. This means that the
inclusion of the mass difference leads to a transition potential that is
generally weaker and of shorter range, a feature which has already been 
observed e.g.~in \cite{hol78}.

If one furthermore employs the non-relativistic limit, one obtains the local
representation of the OPE of (\ref{resultloc}), which is  given in
Fig.~\ref{fig1} by the dashed line (local). The removal of the relativistic
features yields a sizeable effect at larger momenta in particular. The dashed
dotted line (contact rem.) results from the expression of (\ref{resultloc}) if
the contact term, the constant in the momentum representation, is ignored. The
removal of this contact term has a very strong effect on the transition
potential in this central partial wave with $l=0$. Only after removing this
contact term, we obtain an OPE component of the Bonn potential, which is essentially
identical to the OPE contribution in the Argonne V28 model.

\begin{figure}[t]
\begin{center}
\epsfig{figure=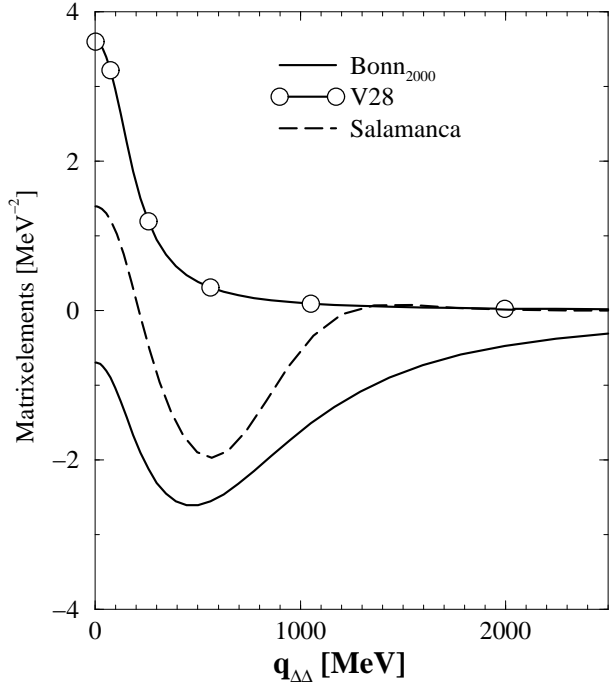,width=8cm}
\end{center}
\caption{Matrix elements for the transition potential $NN \to \Delta\Delta$ in the
$^3S_1$ central partial wave assuming the Bonn$_{2000}$, the Argonne V28 and the
Salamanca quark model approach. Further details see Fig.~\protect{\ref{fig1}}.
\label{fig2}}
\end{figure} 
The comparison of these various approximations in Fig.~\ref{fig1} makes it
rather obvious that the different treatment of the OPE contribution to the $NN
\to \Delta\Delta$ transition potential leads to quite different matrix elements
in the relativistic Bonn model, defined in momentum space, as compared to the
local treatment in the Argonne V28. 

This OPE contribution is the most important ingredient to the
transition potential in both models. The addition of the $\rho$ exchange in the
case of the Bonn$_{2000}$ model leads to minor although non-negligible
modifications. This can be seen from the comparison of Fig.~\ref{fig1} with 
Fig.~\ref{fig2}, which shows matrix elements of the total transition potential.
Besides the results for the Bonn$_{2000}$ and V28 models, this figure also shows
the corresponding values for the quark model of the Salamanca group
\cite{entem}. We see again the huge differences between the V28 and the
Bonn$_{2000}$ model, which even lead to a difference in sign. From our
discussion above we know that the main part of differences can be traced back 
to the different treatment of the OPE contribution. 

The Salamanca Chiral Quark Cluster (CQC) model, which we study as a third 
example\cite{entem} cannot be considered as an entirely realistic potential 
because it does not fit the $P$-phase shifts for NN scattering with good 
accuracy; the phase shifts in the ${}^1S_0$ and ${}^3S_1-{}^3D_1$ 
channels are reproduced very well, however.
In the CQC model, the long-range interaction is generated by OPE as well.
In the core region,
the potential is determined by gluon exchange and the Pauli principle.
The latter requires total antisymmetry of the  
six-quark wave function and this leads to a dependence of the
baryon-baryon potential on the initial kinetic energy $E_{in}$ of two 
interacting nucleons. For the results displayed in Fig.~\ref{fig2} we assume a
value $E_{in}=0$ for this energy. 

The matrix elements for the transition potential of the Salamanca model exhibit
values in between the two approaches discussed above. The shape is similar to
the one of the Bonn potential, however, approaching the value of zero with
increasing momenta much faster than the Bonn potential. This might be an
indication that the Pauli effects in the quark model provide a much stronger
cutoff at high momenta than the form factor used in the Bonn potential. 
\begin{figure}
\begin{center}
\epsfig{figure=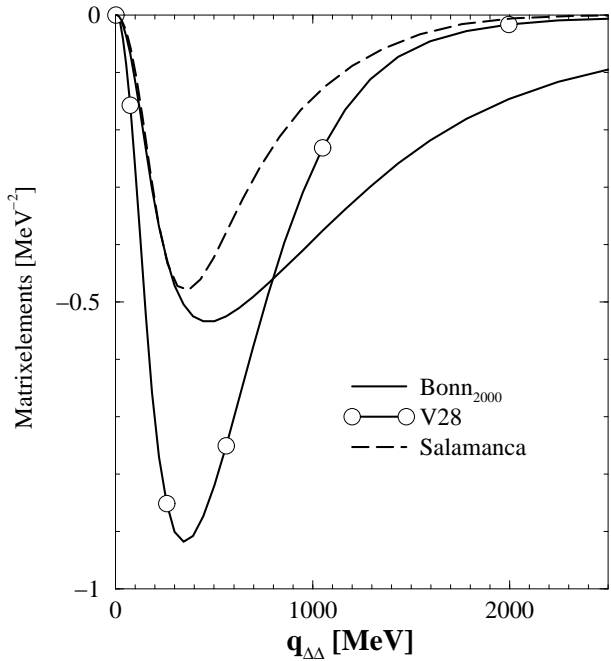,width=8cm}
\end{center}
\caption{Matrix elements for the transition potential
$\big<{q^{\prime}\,{}^3\!S_1^{NN}}\big|{V}\big|
{{q}\,{}^7\!D_1^{\Delta\Delta}}\big>$ assuming the Bonn$_{2000}$, the Argonne
V28 and the Salamanca quark model approach. Further details see
Fig.~\protect{\ref{fig1}}. \label{fig3}}
\end{figure}
As another example we compare results obtained for these three models also for 
the case $\big<{q^{\prime}\,{}^3\!S_1^{NN}}\big|{V}\big|
{{q}\,{}^7\!D_1^{\Delta\Delta}}\big>$ in Fig.~\ref{fig3}. The differences
between Bonn$_{2000}$ and V28 can again be traced back to the
different treatment of the OPE. For this partial wave, however, one does not
obtain any contact term contribution. This implies that the V28 is stronger at
small momenta as it ignores the mass difference $m_\Delta - m_N$ in the pion
propagator and weaker than Bonn$_{2000}$ at high momenta because of the
non-localities included in the Bonn model. The matrix elements for the Salamanca
model are similar to those of the Bonn$_{2000}$ potential at small momenta but
exhibit features of a stronger cut-off at high momenta. 

\section{The wave functions of the deuteron}
\label{The wave functions of the deuteron}
Solutions of the bound state two-particle problem 
have been obtained for the Argonne V28, the $\mbox{Bonn}_{2000}$
and the Salamanca CQC potential. The nucleonic part of the deuteron wave
function contains components ${}^3S_1-{}^3D_1$ partial waves.
If $\Delta$ degrees of freedom shall be taken into account explicitely,
one has to extend the two coupled nucleonic channels
by four $\Delta\Delta$ partial waves, namely
${}^3S_1^{\Delta\Delta}$,
${}^3D_1^{\Delta\Delta}$,
${}^7D_1^{\Delta\Delta}$ and
${}^7G_1^{\Delta\Delta}$.
No $N\Delta$ states can occur because the deuteron is an isospin
$T=0$ state and the isospins of $N$ and $\Delta$ cannot
couple to zero.
The problem to determine the baryonic wave function in the 6 coupled channels
has been solved in a spherical box with radius $R_{box}=20\,\mbox{fm}$. The
spherical Bessel functions with the boundary condition
$$j_l(k_{il}R_{box}) = 0$$
can be used to construct a complete basis of orthonormalized states for the
bound state wave functions within the spherical box\cite{box}.  The coupled
channel Hamiltonian was calculated in this basis of  momentum eigenstates of
the box and diagonalised numerically. Up to 300 discrete momenta were needed
to get stable results for the binding energy. The results for energies and the
wave functions obtained with the  Argonne V28 and the Salamanca potential   are
in good agreement with the values given  in~\cite{V28} and~\cite{entem},
respectively.

All $NN$ interactions, which we consider, are adjusted to fit the total energy of
the deuteron, which can be written as a sum of the kinetic energy $T_{total}$
and potential energy $V_{total}$ with
\begin{eqnarray}
T_{total} & = & \left< NN\,^3S_1| T | NN\,^3S_1 \right> +  \left< NN\,^3D_1| T |
NN\,^3D_1 \right> + \left< \Delta\Delta | T | \Delta\Delta \right> \nonumber \\
& = & T^N_S + T^N_D + T^\Delta\,. \label{eq:tkin}
\end{eqnarray}
Note that the term $T^\Delta$ sums up the kinetic energy from all
partial wave components in the $\Delta\Delta$ wave function and includes the
contribution from the $N - \Delta$ mass difference. The potential energy can be
split into
\begin{eqnarray}
V_{total} & = & \left< NN\,^3S_1| V | NN\,^3S_1 \right> +  \left< NN\,^3D_1| V |
NN\,^3D_1 \right> + \nonumber \\ && \qquad 
2\left< NN\,^3S_1| V | NN\,^3D_1 \right> + 
2\left< NN | V | \Delta\Delta \right> +\left< \Delta\Delta | V | \Delta\Delta 
\right>\nonumber \\
& = & V^{NN}_{SS} + V^{NN}_{DD}  +  V^{NN}_{SD} +  V^{N\Delta} +
V^{\Delta\Delta}\,. \label{eq:vpot}
\end{eqnarray}
\begin{table}[t]
\begin{center}
\begin{tabular}{c|rrr|rr} 
 & Arg. V28 & Bonn$_{2000}$ & Sal. CQC $\quad$& Arg. V14 & CD-Bonn\\ 
\hline\hline
$T_{total}$   & $23.75 $ & $22.29$ & $17.82\quad$ & $19.22$ & $15.48$\\  
$V_{total}$   &  $-25.97$  & $-24.51$ & $-20.04\quad$ & $-21.44$ & $-17.70$\\
\cline{1-6}
$T_S^{{N}}$ & $10.32$ & $10.25$& $10.84\quad$ & $10.54$ & $9.79$ \\
$T_D^{{N}}$ & $8.95$ & $5.35$ & $5.13\quad$ & $8.68$ & $5.69$ \\
$T^{\Delta}$ & $4.48$ & $6.69$ & $1.85\quad$ & - & - \\
\cline{1-6}
$V_{SS}^{{NN}}$ & $7.08$ & $1.52$ & $-5.67\quad$ & $-1.83$ & $-4.77$ \\
$V_{DD}^{{NN}}$& $5.86$ & $1.80$ & $0.70\quad$  & $1.99$ & $1.34$ \\
$V_{SD}^{{NN}}$ & $-29.22$ & $-14.38$ & $-11.64\quad$ & $-21.60$  & $-14.27$ \\
$V^{N\Delta}$ & $-10.40$ & $-13.44$ & $-3.16\quad$ & - & - \\ 
$V^{\Delta\Delta}$& $0.71$ & - & $-0.28\quad$  & - & - \\
\end{tabular}
\caption{\label{deuter1}
Contributions to the kinetic and potential energy of the deuteron
originating from the different parts of the wave function. The kinetic energy
$T_{total}$ is the sum of the kinetic energies originating from the $NN$ wave
function in $^3S_1$ ($T_S^{N}$) and $^3D_1$ ($T_D^{N}$) partial waves plus the
kinetic energy ($T^\Delta$) from the $\Delta\Delta$ components. The term
$T^\Delta$ also accounts for the $N\Delta$ mass difference. The potential energy
contains contributions from the various parts of the $NN \to NN$ potential
($V^{NN}$), the $NN \to \Delta\Delta$ terms ($V^{N\Delta}$) and the
$\Delta\Delta \to \Delta\Delta$ terms ($V^{\Delta\Delta}$). Results are given
for the 3 models with isobar configurations. For comparison we also show
results
from pure nucleonic interaction models V14 and
CDBonn. All entries are given in MeV.}
\end{center}
\end{table}
The contributions of these various terms to the energy of the deuteron are
listed in table \ref{deuter1}, while the norm of the various partial wave
contributions to the wave function are presented in table \ref{deuter2} for the
three interaction models with inclusion of isobars. For a comparison table
\ref{deuter1} also shows the result for two NN interaction models without the
explicit treatment of isobar configurations: the Argonne potential model V14
\cite{V28} and the charge-dependent Bonn potential \cite{cdb}.

\begin{table}[t]
\begin{center}
\begin{tabular}{c|ccc}
\multicolumn{1}{c}{\%} & \multicolumn{1}{c}{$\mbox{V}_{28}$} & 
\multicolumn{1}{c}{$\mbox{Bonn}_{2000}$} & \multicolumn{1}{c}{CQC}
 \\ \hline\hline
${}^3S_1^{{NN}}$ & 93.341 & 94.685 & 95.199 \\
${}^3D_1^{{NN}}$ & 6.133 & 4.705 & 4.567 \\
${}^3S_1^{\Delta\Delta}$ & 0.043 & 0.223 & 0.107 \\
${}^3D_1^{\Delta\Delta}$ & 0.020 & 0.022 & 0.004 \\
${}^7D_1^{\Delta\Delta}$ & 0.417 & 0.361 & 0.124 \\
${}^7G_1^{\Delta\Delta}$ & 0.045 & 0.005 & 0.006 \\
Total $\Delta\Delta$ & 0.524 & 0.611 & 0.241 \\ 
\end{tabular}
\caption{\label{deuter2}Probability for D-state and 
$\Delta\Delta$-components in the deuteron for different models of the baryon -
baryon interaction. All entries in percent.
}
\end{center}
\end{table}
Although the sum of these various energy contributions yields the same
energy of the deuteron for all the interaction models, there are remarkable
differences in the individual terms. The inclusion of isobar
components in the wave function yields a contribution to the binding energy of -5.21 MeV, -6.75 MeV
and -1.59 MeV for the V28, the Bonn$_{2000}$ and the Salamanca interaction,
respectively. This means that V28 and Bonn$_{2000}$ predict an unbound deuteron,
if the $\Delta\Delta$ components in the wave function would be suppressed, while
the Salamanca interaction leads to a much weaker contribution of the isobar
configurations. 

\begin{figure}[t]
\begin{center}
\epsfig{figure=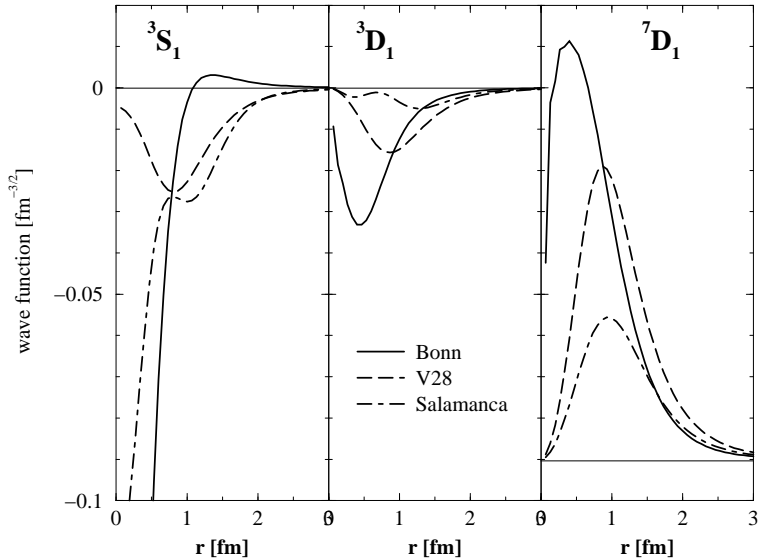,width=10cm}
\end{center}
\caption{Isobar components in the deuteron wave function using the three
different models for the baryon-baryon interaction. \label{fig4}}
\end{figure} 
These differences also show up in the norm of the various partial
wave components in the deuteron wave function listed in table \ref{deuter2}.
Comparing these occupation probabilities one finds that all interactions predict
the largest contribution to the $\Delta\Delta$ probability in the $^7D_1$
partial wave. The largest probability for this partial wave is obtained for the
V28 potential, while Bonn$_{2000}$ and the Salamanca interaction in particular
yield much smaller probabilities in this channel. The origin of these differences
can be deduced from Fig.~\ref{fig4}, which exhibits the $\Delta\Delta$ wave
functions. The Argonne potential leads to a wave function in the $^7D_1$,
$\Delta\Delta$ channel, which is of longer range than those derived from the
other two interactions. The main reason for this feature is the
fact that the OPE contribution to the $NN \to \Delta\Delta$ transition potential
neglects the $N - \Delta$ mass difference in the pion propagator (see (\ref{feyn})),
which leads to a transition potential of longer range.
Comparing the wave functions in this channel derived from the Bonn$_{2000}$ and
the Salamanca interaction, one observes that the latter is strongly suppressed
at short relative distances as compared to the former, which supports the
argument presented already in the comparison of the matrix elements of the
interaction, that the quark model leads to a much stronger effective cut-off 
at short distances than the cut-offs used in the meson exchange model.

The most significant difference in the predictions for the isobar components of
the wave function can be seen in the $^3S_1$, $\Delta\Delta$ channel. While the
Bonn and Salamanca models yield occupation probabilities, which are of similar
size as those for the $^7D_1$, $\Delta\Delta$ channel, the probability for this
channel derived from the V28 potential is much weaker. A qualitative difference
can also be observed from the inspection of the wave function in this channel
(see Fig.~\ref{fig4}). The wave function obtained for the Bonn and Salamanca
potential show a maximal amplitude for $r \to 0$, while the corresponding wave
function deduced from the V28 potential is suppressed at small relative
distances. These differences in the wave function reflect the fact that the
contact term in the OPE contribution to the transition potential (see
(\ref{resultloc}) and discussion there) is removed in the local representation
of the OPE in V28, while the non-local representation of the OPE in the other
models keeps a strong short range component.

The comparison of the $\Delta\Delta$ components in the deuteron wave function
therefore reflects the main differences between the local approximation of the
V28 and the non-local approaches of the Bonn and Salamanca model: The V28
contains a OPE component of longer range and suppresses the short range
components much stronger than the other two models. Furthermore, the quark model
approach leads to a stronger reduction than the typical cut-off that is
used in the
meson exchange models of the Bonn group. This leads to much weaker
$\Delta\Delta$ components in the deuteron wave function for the Salamanca model.

It is worth noting that the contributions of $V_{SD}^{NN}$ to the potential
energy of the deuteron (see table \ref{deuter1}) are significantly larger for
the local Argonne potentials V14 and V28, than for the various versions of the
Bonn potential and the Salamanca potential. This observation suggests that the
local representation of the tensor components of the NN interaction, in
particular the contribution originating from $\pi$-exchange yields matrix
elements which are significantly larger than those evaluated within the non-local approaches (see also \cite{local,poldeut}). 
\section{Correlations in nuclear matter and ${}^{16}\mbox{O}$}
\label{Two-body correlations}
The wave functions of many-body systems are studied in the
framework of the coupled-cluster theory~\cite{kuem}.
In the coupled-cluster approach  one starts assuming
an appropriate Slater determinant $\Phi$ and writes the exact
eigenstate $\Psi$ for the $A$-particle system as
\begin{equation}
\Psi = e^S \Phi \label{eq:esansatz}\mbox{,}
\end{equation}
with $S$ an operator of the form
\begin{equation}
S = \sum_{n=1}^A S_n \label{eq:sumsn}\mbox{,}
\end{equation}
where $S_n$ is an $n$-particle operator which can be written for the case of
$n=2$
\begin{equation}
S_2 = \frac{1}{4}\sum_{\nu_1,\nu_2, \rho_1, \rho_2}
\left<\rho_1 \rho_2 \vert S_2 \vert \nu_1\nu_2 \right> a^\dagger_{\rho_1} 
a^\dagger_{\rho_2} a_{\nu_2}  a_{\nu_1} \label{eq:sndef}\mbox{.}
\end{equation}
In this equation $a^\dagger_{\rho_i}$ stand for fermion creation operators in
states which are unoccupied in $\Phi$, while  $a_{\nu_i}$ represent
annihilation operators for the nucleon single-particle states which are occupied
in the Slater determinant $\Phi$. Note that the $a^\dagger_{\rho_i}$ may also
represent the creation of $\Delta$ isobar states. Therefore the $S_2$ amplitudes
describe two-particle two-hole excitations relative to $\Phi$ but also $N\Delta$
and $\Delta\Delta$ excitations.

For our present investigation of nuclear matter we will assume $\Phi$ to be 
the Slater determinant defined in terms of plane waves, occupying all states
with momenta up to the Fermi momentum $k_F$ = 1.36 fm$^{-1}$. As an example for
a finite nucleus, we will also consider $^{16}$O. In this case we will assume
$\Phi$ to be the Slater determinant, defined in terms of harmonic oscillator
states ($\hbar\omega$ = 14 MeV) with nucleons occupying the states of $0s$ and
the $0p$ shell. If one assumes that these single-particle states represent an
optimal single-particle basis, i.e.~the amplitudes $S_1$ in (\ref{eq:sumsn})
vanish, and ignores the contributions of linked $n$-particle correlations with
$n\geq 3$ ($S_n = 0$ for $n\geq 3$), one obtains integral equations for
amplitudes
\begin{equation}
\left< b_1b_2\left[k(lS)j\right] KLJ\tau \vert S_2 \vert (\nu_1\nu_2) J\tau
\right> \label{def:s2amp}\mbox{,}
\end{equation}
which are solved in momentum space\cite{stauf}. In this representation $b_1b_2$
stand for $NN$, $N\Delta$ and $\Delta\Delta$ states, $k(lS)j$ denote the
momentum, spin and orbital quantum numbers for the partial wave basis of the
relative motion of the two baryons. $K$ and $L$ represent the center of mass
state and $J$ and $\tau$ refers to the total angular momentum and isospin of the
pair of baryons. 

The $S_2$ amplitudes can be considered as correlation functions describing the
difference between the correlated and uncorrelated wave function of
two particles in the nuclear medium. As the uncorrelated Slater determinant
$\Phi$ in (\ref{eq:esansatz}) does not contain any isobar components, the $S_2$
amplitudes can be interpreted directly as the $N\Delta$ and $\Delta\Delta$
component of the two-particle wave function, if $b_1b_2$ in (\ref{def:s2amp})
refer to $N\Delta$ and $\Delta\Delta$ states.

\begin{figure}[t]
\begin{center}
\epsfig{figure=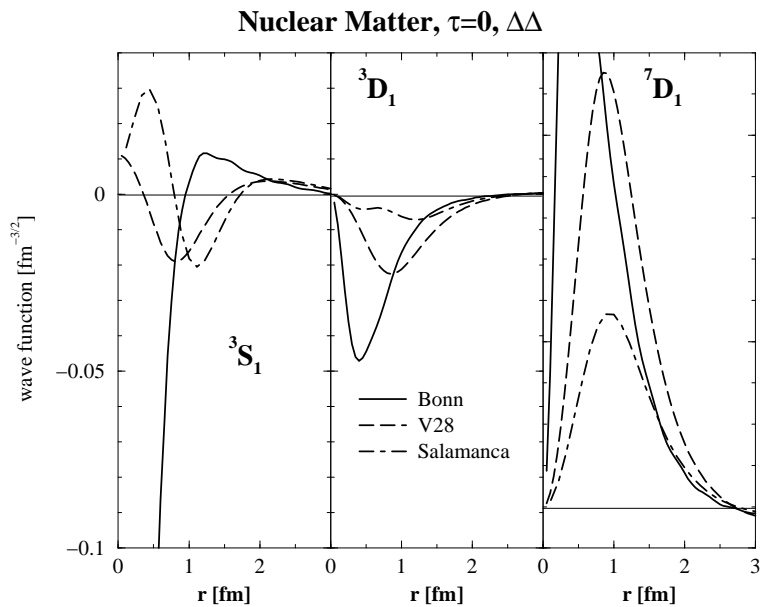,width=10cm}
\end{center}
\caption{Isobar components in the relative wave functions of two particles in
nuclear matter with Isospin $\tau = 0$ and $J=1$. \label{fig5}}
\end{figure}
\begin{figure}[t]
\begin{center}
\epsfig{figure=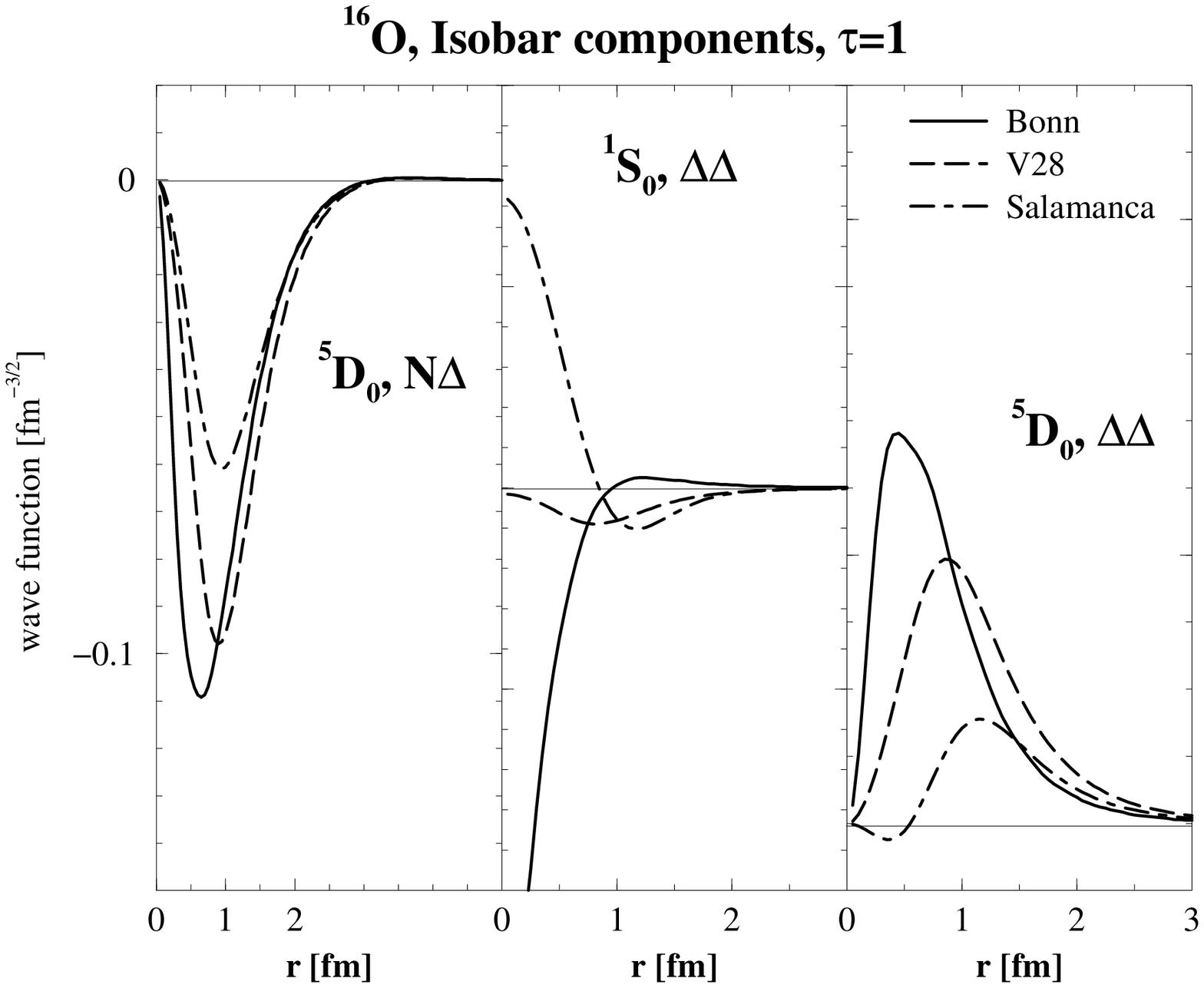,width=10cm}
\end{center}
\caption{Isobar components in the relative wave functions of two particles in
$^{16}$O with Isospin $\tau = 1$ and $J=0$. The uncorrelated state
corresponds to two nucleons in the $0s_{1/2}$ state.\label{fig6}}
\end{figure}
As typical examples for these correlation functions in nuclear matter, the
$\Delta\Delta$ components for the relative wave function of two baryons in a
state with isospin $\tau = 0$ and angular momentum $j=1$ are given in
Fig.~\ref{fig5}. These components correspond to the components of the deuteron
wave function in the same partial waves displayed in Fig.~\ref{fig4}. In fact,
these isobar components of the wave function for a pair of baryons in nuclear
matter is very similar to the deuteron wave function if one compares the
partial wave with $l=2$. The $\Delta\Delta$ wave functions exhibit a tail of longer
range in the case of the Argonne V28 interaction, which reflects the longer
range of the $NN \to \Delta\Delta$ transition potential for this interaction as
compared to the other two approaches. The Salamanca CQC approach yields
amplitudes for these $D$-waves which are considerably smaller. All these
amplitudes are enhanced as compared to the deuteron wave function, which
reflects the larger density of the nuclear matter system.

The situation is a little bit different if one compares the $^3S_1$,
$\Delta\Delta$ component of the nuclear matter wave function with the
corresponding part of the deuteron wave function. The results obtained for the
V28 and the Bonn potential yield wave functions of rather similar shape in the
deuteron and in nuclear matter also for this channel. The wave functions 
derived from the Salamanca CQC interaction 
are rather different in nuclear matter as
compared to the deuteron. At short distances they even have a different sign.
This reflects a large non-locality or momentum dependence of the short range
component for the  $NN \to \Delta\Delta$ transition potential derived in the CQC
model. 

In the nuclear many body systems one also observes bound states of two baryons
in states with isospin $\tau =1$. As an example of such configurations we
discuss $N\Delta$ and $\Delta\Delta$ components of two-particle wave functions
in ${}^{16}\mbox{O}$ as displayed in Fig.~\ref{fig6}. 
The comparison of the results
obtained for the different interaction models leads to observations which are
rather similar to those discussed for the states with $\tau=0$ above. The difference in
the range of the transition potentials $NN \to N\Delta$ for V28 and the other
two interactions is not as significant as in the case of the $NN \to 
\Delta\Delta$ (compare the discussion of the propagator in (\ref{feyn})), which
leads to smaller differences in the tail of the $^5D_0$ $N\Delta$ wave function 
than in the corresponding $\Delta\Delta$ state. A very significant model
dependence is obtained in the $^1S_0$, $\Delta\Delta$ wave function. 

The isobar-isobar relative wave functions derived for nuclear matter and finite nuclei are
rather similar. Therefore we display only one of these examples for each
channel. This demonstrates that isobar admixtures correspond to
correlations in the many-body wave function which are of short range. Therefore
they are not very sensitive to surface effects in finite nuclei and a local
density approximation should be appropriate to consider isobar effects in finite
nuclei. 
\section{Conclusions}

Three different models for the baryon-baryon interaction, which fit NN
scattering data and explicitely account for isobar degrees of freedom have been
considered. These are the local Argonne V28 potential\cite{V28}, the 
Bonn$_{2000}$\cite{mac00}
interaction model based on the relativistic meson exchange model and an
interaction based on the Chiral Quark Cluster (CQC) model which has recently
been developed by the Salamanca group~\cite{entem}. The isobar components in the
wave function of the deuteron and nuclear many-body systems including nuclear
matter and $^{16}$O as an example for a finite nucleus are evaluated.
Significant differences are observed in the predictions from these models.
These differences can be related to the assumptions made in determining the
transition potential between $NN$, $N\Delta$ and $\Delta\Delta$ states. The
V28 interaction model yields isobar wave functions of longer range than the
other two. This can be traced back to the approximations which are made in reducing
the $\pi$ exchange contribution of the transition potentials to a local form.
The most significant differences are observed in partial waves with $l=0$ at
small distances. A main source of the discrepancies in these partial waves is
related to a removal of a contact term in the local interaction model. The
Bonn$_{2000}$ and Salamanca CQC model in particular exhibit important non-local
features for the short range part of the interaction model. The Salamanca model
predicts rather small isobar components in the nuclear many-body wave function.
These different predictions for the isobar components in the many-body wave
function could be very useful in distinguishing between different interaction
models.

We thank D. Entem for providing us with the Salamanca CQC potential. Also we
would like to thank him but also Francesca Samarucca and Ruprecht Machleidt 
for useful discussions. One of the authors (T.F.) would also like to thank the
Dr.-Carl-Duisberg-Stiftung for supporting this work by a grant that enabled a
stay at the  Departament d'Estructura i Constituents de la Mat\`eria in
Barcelona, Spain.
We would like also to acknowledge financial support from DGICYT (Spain)
under contract PB98-1247 and from Generalitat de Catalunya under grant
SGR2000-24.

\end{document}